# An alternative mathematical theory of elastoplastic behaviour

J. J. Nader

Department of Structural and Geotechnical Engineering, Polytechnic School, University of São Paulo, São Paulo, Brazil.

*Abstract*: This paper presents a theory for the behaviour of isotropic-hardening/softening elastoplastic materials that do not have a preferred reference configuration. In spite of important differences, many ingredients of classical plasticity are present.

Main features are: the elastic and plastic responses are given by solutions of hypoelastic differential equations, no decomposition of the deformation into elastic and plastic parts is done from the start, the hardening rule is an outcome, and the principle of material objectivity is obeyed. An important result is the existence of a limit surface that divides the stress space into regions of hardening and softening and is composed of equilibrium points of the differential equation of plastic response.

Keywords: elastoplastic material, rate-type objective plasticity, hardening/softening plasticity.

## 1. INTRODUCTION

The theory presented in this paper differs from the modern theories of finite plasticity that arose in the last forty years (see, e.g. [1]) but has a similar objective, namely, to arrive at a mathematically clear and physically sound description of elastoplastic behaviour and to enlarge the range of applicability of the classical theory to large deformation problems.

It is known that classical plasticity lacks precision and obedience to the principle of material objectivity. Besides, in most expositions of classical plasticity, there is no clear definition of which stress, deformation tensor and rate-of-deformation tensor is being used; time derivatives are treated as increments, and hypotheses are sometimes tacitly assumed (cf. comments in [2] and [3]). These defects have been a motivation to development of the proposed theory.

The constitutive equations presented in this paper describe the behaviour of isotropic-hardening/softening elastoplastic materials. The theory is organised in a relatively simple manner, with a small number of axioms and constitutive functions. In spite of important differences, many ingredients of classical plasticity are present.

In what follows we state the main features of the theory.

a) The principle of material objectivity is obeyed. This theory is, therefore, in principle appropriate for large deformation problems.
b) The constitutive equations involve only quantities defined in the current configuration: the Cauchy stress $T$ and the spatial velocity gradient $L$. Both the elastic and plastic responses are given by solutions of hypoelastic differential equations.
c) The problem of determining the material response (the Cauchy stress) arising in a given motion is treated with full generality. The way classical and some finite plasticity theories are usually constructed, this problem is not treated as simply as in this theory. Traditionally, the plastic part of the deformation is unknown and is related (through the so-called flow rule) to the stress and to the stress rate, which are also unknown, when the motion is given.
d) The constitutive equations for elastic and plastic responses are hypoelastic differential equations and as such they must be integrated in intervals. The yield criterion is established in intervals, as it must be since the equations involve derivatives, in contrast to the usual approach to rate-type plasticity, in which the yield criterion and the equations refer to isolated instants. Traditionally, the constitutive equations are treated as if they involved increments of stress and strain rather than time derivatives, and this hides general properties of solutions.

e) Here there is no axiomatic decomposition of the deformation into elastic and plastic parts. In rate-type theories of plasticity the decomposition of the time derivative of the infinitesimal strain tensor (in classical plasticity) or of the stretching tensor (in modern theories) is imposed at the outset, but is rather artificial because, although it is true that in a cycle of stress the body does not end up in the same original configuration due to irreversible strains, it is not natural that at each stage of the process the rate of deformation should be seen as the sum of a reversible and an irreversible part. Recall that the two parts of the stretching are not the symmetric part of the velocity gradient of any motion as its sum is. In this work no decomposition is necessary. However, to make the comparison with other rate-type theories, we will show how the stretching tensor can be considered as the sum of elastic and plastic parts.

f) The yield surface is defined in stress space and the yield criterion involves both the Cauchy stress and the stretching tensor.

g) No hardening rule is set from the start. It is obtained as an outcome.

h) The theory incorporates hardening, softening and perfectly plastic behaviour.

i) An interesting result is that there exists a limit surface that divides the stress space into regions of hardening and softening. This surface is composed of equilibrium points of the differential equation of plastic response for certain motions.

Characteristic (b) is important if one wishes to model the behaviour of materials that have no preferred reference configurations, like soils. For this kind of material it seems to be more consistent to work with stress and rate-of-deformation tensors defined in the current configuration; we wish to know how the stress will change, from the current state, if a certain motion is imposed. The motive is the same that led Truesdell [4] to propose hypoelasticity. Property (f) is also compatible with materials that have no preferred configurations. For those materials it would not make sense to define the yield surface in strain space.

Property (i) deserves special attention (sections 3 and 4). It allows modelling materials whose failure surface is different in form from the yield surface. It is especially interesting to model critical states of soils.

Throughout the paper the standard notation of modern continuum mechanics is employed. Frequently used symbols are: *R* (the set of all real numbers), *Lin* (the set of all second order tensors), *Lin*$^+$ (the set of all elements of *Lin* with positive determinant), *Sym* (the set of all symmetric elements of *Lin*), *Skw* (the set of all skew elements of *Lin*), *Orth* (the set of all orthogonal elements of *Lin*). The transpose of a tensor $A$ is denoted $A^T$; $A$:$B$ indicates the inner product of $A$ and $B$ defined as $A$:$B$=$tr(A^T B)$, where *tr* is the linear functional trace, and $\|A\|$ denotes the Euclidean norm of $A$. A dot superposed to a symbol denotes material time derivative.

## 2. CONSTITUTIVE EQUATIONS AND MATERIAL RESPONSE

Let the deformation gradient at a point of a body, at each instant *t* of an interval *I*, be given by a piecewise continuously differentiable function[1] $F$:$I{\subset}R{\rightarrow}Lin^+$.

The response of the elastoplastic material subject to the deformation gradient function $F$ is represented by $T$:$I{\rightarrow}Sym$, which associates to each instant *t* the Cauchy stress $T(t)$. The functions $F$ and $T$ are linked by hypoelastic differential constitutive equations, as we shall see. In the description of the material response there will appear the function $k$:$I{\rightarrow}R$, which associates to each instant the hardening parameter $k(t)$.

The spatial velocity gradient $L = \dot{F}F^{-1}$ will enter into the constitutive equations through its symmetric part, the stretching tensor $D$, and its skew-symmetric part, the spin tensor $\boldsymbol{W}$. The corotational or Jaumann rate of $\boldsymbol{T}$, defined by $\overset{\circ}{\boldsymbol{T}} = \dot{\boldsymbol{T}} - \boldsymbol{W}\boldsymbol{T} + \boldsymbol{T}\boldsymbol{W}$, will also be used.

---

[1] Thus, there may exist instants in *I* at which the left-hand derivative of $F$ does not equal the right-hand derivative. It is important to consider this possibility, rather than to restrict the analysis to a continuously differentiable $F$, because in plasticity abrupt changes in the direction of deformation, like strain reversals, must be taken into account.

The definition of the material response will refer to closed sub-intervals of $I$ on which the restriction of $\boldsymbol{F}$ is continuously differentiable. In this respect, it is important to state clearly a convention regarding derivatives at the extremes of a closed interval. Suppose that $\boldsymbol{F}$ fails to be continuously differentiable at $a$ and $b$ but is continuously differentiable on $[a,b]$. In studying the material response in the interval $[a,b]$, by $\dot{\boldsymbol{F}}(a)$ and $\dot{\boldsymbol{F}}(b)$ we mean the right-hand derivative of $\boldsymbol{F}$ at $a$ and the left-hand derivative of $\boldsymbol{F}$ at $b$, respectively. The same convention applies to $\boldsymbol{T}$ and other functions calculated from $\boldsymbol{F}$.

The elastoplastic material is characterised by three constitutive functions:

$$\boldsymbol{A}: Sym \rightarrow L(Sym),$$
$$\boldsymbol{B}: Sym \rightarrow Sym,$$
$$f: Sym \rightarrow R.$$

The symbol $L(Sym)$ stands for the set of all linear operators of $Sym$ (a fourth-order tensor space).

The functions $\boldsymbol{A}$, $\boldsymbol{B}$ and $f$ will take part in the differential constitutive equations and have such smoothness properties so as to ensure existence and uniqueness of solution of initial value problems. The principle of material objectivity places restrictions upon these functions, as we shall see.

At each $t \in I$ we define the elastic domain as the set

$$E(t) = \{\boldsymbol{T} \in Sym \,/\, f(\boldsymbol{T}) \leq k(t)\}$$

and its boundary,

$$\partial E(t) = \{\boldsymbol{T} \in Sym \,/\, f(\boldsymbol{T}) = k(t)\},$$

is the yield surface.

As an axiom, we require that

$$T(t) \in E(t), \forall t \in I. \qquad (1)$$

*Material Response*

Let the restriction of $F$ to $[a,b] \subset I$ be continuously differentiable. Suppose that at $t_0 \in [a,b)$ the stress is $T_0$, the hardening parameter is $k_0$, and that they are such as to satisfy the axiom established in (1). We will now specify the material response $t \mapsto T(t)$, along with the function $t \mapsto k(t)$, on sub-intervals $[t_0, \tau]$, $a \le t_0 < \tau \le b$.

For the sake of simplicity we introduce $\psi(T,D) = A(T)[D] : \nabla f(T)$ and $D_0 = D(t_0)$. The symbol $\nabla f(T)$ denotes the gradient of $f$ at $T$.

To characterise the material response consider four possible initial situations.

Case I: $f(T_0) < k_0$.
Case II: $f(T_0) = k_0$ and $\psi(T_0, D_0) < 0$.
Case III: $f(T_0) = k_0$ and $\psi(T_0, D_0) = 0$.
Case IV: $f(T_0) = k_0$ and $\psi(T_0, D_0) > 0$.

a) In cases I, II and III, the response is given by $T$ satisfying $T(t_0) = T_0$ and

$$\overset{\circ}{T} = A(T)[D] \qquad (2)$$

in any interval $[t_0, \tau]$ such that $f(T(t)) \le k_0$, $\forall t \in [t_0, \tau]$. In such intervals the response is said to be elastic and the hardening parameter is constant:

$$k(t) = k_0.$$

b) In cases III and IV, the response is given by $T$ satisfying $T(t_0)=T_0$ and

$$\overset{\circ}{T} = A(T)[D] + \psi(T,D)B(T) \qquad (3)$$

in any interval $[t_0,\tau]$, such that $\psi(T(t),D(t))>0$, $\forall t \in (t_0,\tau)$. In these intervals the response is said to be plastic and the hardening parameter varies according to

$$k(t)=f(T(t)) \qquad (4)$$

The function $B$ appearing in (3) is such that $B(T) \neq 0$, for all $T$.

Proposition 1 below shows that this definition of material response is complete and definite in the sense that all possible situations are taken into account and that the response cannot be simultaneously elastic and plastic. In short, after $t_0$ either one type of response or the other will follow. This is a minimum requirement that a reasonable theory should satisfy.

*Proposition 1*: In cases I and II intervals as specified in (a) exist. In case III there exist either intervals as specified in (a) or intervals as specified in (b). In case IV intervals as specified in (b) exist.

*Proof*: That the specified intervals exist in cases I and IV is clear from the continuity of the functions involved.

In cases II and III, intervals as specified in (a) exist if and only if $f \circ T$ is a non-increasing function in a right-neighbourhood of $t_0$. This certainly happens in case II, but it may not happen in case III, as is shown next.

Let us compute the derivative of $f \circ T$ in a right-neighbourhood of $t_0$, when the response is elastic (2):

$$\frac{d}{dt}(f \circ T)(t) = \nabla f(T(t)) : \dot{T}(t) = \nabla f(T(t)) : A(T(t))[D(t)] = \psi(T(t),D(t)).$$

Thus, in case II, $f \circ \boldsymbol{T}$ decreases in a right-neighbourhood of $t_0$, since $\psi(\boldsymbol{T}_0, \boldsymbol{D}_0)<0$. In case III, as $\psi(\boldsymbol{T}_0, \boldsymbol{D}_0)=0$, nothing can be said about the increase of $f \circ \boldsymbol{T}$.

According to the definition of material response, case III can be the beginning of elastic or plastic response. It will now be shown that case III gives rise either to elastic or to plastic response.

If, considering the solution of (2) in (a), $t \mapsto \psi(\boldsymbol{T}(t),\boldsymbol{D}(t))$ does not increase (remains 0 or becomes negative) in a right-neighbourhood of $t_0$, then the response is elastic in a right-neighbourhood of $t_0$, for $f \circ \boldsymbol{T}$ will not increase as well.

If, considering the solution of (3) in (b), $t \mapsto \psi(\boldsymbol{T}(t),\boldsymbol{D}(t))$ increases (becomes positive) in a right-neighbourhood of $t_0$, then the response is plastic in that neighbourhood.

The increase of $t \mapsto \psi(\boldsymbol{T}(t),\boldsymbol{D}(t))$ near $t_0$ will now be studied by means of its derivative. Let us introduce $\boldsymbol{P}(\boldsymbol{T})=\boldsymbol{A}(\boldsymbol{T})^T[\nabla f(\boldsymbol{T})]$, so that we can write $\psi(\boldsymbol{T},\boldsymbol{D})=\boldsymbol{D}:\boldsymbol{P}(\boldsymbol{T})$, and compute the following derivative:

$$\frac{d}{dt}\psi(\boldsymbol{T}(t),\boldsymbol{D}(t)) = \dot{\boldsymbol{D}}(t):\boldsymbol{P}(\boldsymbol{T}(t)) + \boldsymbol{D}(t):\mathrm{d}\boldsymbol{P}(\boldsymbol{T}(t))[\dot{\boldsymbol{T}}(t)],$$

in which $\mathrm{d}\boldsymbol{P}$ indicates the Fréchet-derivative of $\boldsymbol{P}$. As both equations (2) and (3) give, in this case, the same value for $\dot{\boldsymbol{T}}(t_0)$, then, by the above equation, $\frac{d}{dt}\psi(\boldsymbol{T}(t_0),\boldsymbol{D}(t_0))$ is the same no matter the solution of (2) or (3) is used. If it is negative, the response is elastic; if it is positive, the response is plastic. If it is zero, the second-order derivative must be examined and so on. Therefore, one and only one response takes place. (*End of the proof.*)

From the definition of the material response we can draw some conclusions about the occurrence of cases I to IV.

Suppose first that $t_0 \in (a,b)$. Recall that in $(a,b)$ the functions $\boldsymbol{T}$, $f \circ \boldsymbol{T}$ and $\psi(\boldsymbol{T}, \boldsymbol{D})$ are continuous and, in a sub-interval in which the response is elastic or plastic, $\boldsymbol{T}$ and $f \circ \boldsymbol{T}$ have a continuous derivative.

If $t_0$ is an interior point of a sub-interval of elastic response, then either case I or III ($t_0$ is a point of maximum of $f \circ T$) holds.

If $t_0$ is an interior point of a sub-interval of plastic response, then case IV holds.

If $t_0$ is an instant of transition from elastic to plastic response, then either case III or case IV holds.

If $t_0$ is an instant of transition from plastic to elastic response, then case III holds. Note that, in the characterisation of the plastic response, $\psi(T, D)$ is allowed to be 0 at the extremes of a closed interval and so a transition from plastic to elastic response is possible at a point where $D$ and, hence, $\psi(T,D)$ are continuous.

Suppose now that $t_0=a$ and that $F$ is not continuosly differentiable at $a$. Then, any of the four cases may happen (at $t_0=a$, $D$ and, hence, $\psi(T,D)$ have different limits from the left and from the right of $t_0$).

*Restrictions imposed by the principle of material objectivity upon the constitutive functions*

The fact that a certain $T$ belongs or not the elastic domain should not depend on the observer. So, we impose that $f(QTQ^T)=f(T)$, for all $T \in Sym$ and for all $Q \in Orth$. Hence, the yield function must be isotropic. It can be shown that the gradient of the yield function, $\nabla f$, is isotropic too: $\nabla f(QTQ^T)= Q\nabla f(T)Q^T$.

The function $(T,D) \mapsto A(T)[D]$ must be isotropic in order that (2) be objective: $A(QTQ^T)[QTQ^T]=QA(T)Q^T$, for all $T$, $D \in Sym$ and for all $Q \in Orth$. Note that $(T,D) \mapsto \psi(T,D)$ results objective.

In order for (3) to be objective, the function $B$ must also be isotropic: $B(QTQ^T)=QB(T)Q^T$, for all $T$, $D \in Sym$ and for all $Q \in Orth$.

## 3. THE EVOLUTION OF THE HARDENING PARAMETER AND THE LIMIT SURFACE

We can obtain more information on how *k* changes in plastic response differentiating both members of (4) with respect to time and using (3):

$$\dot{k} = \frac{d}{dt}(f \circ \boldsymbol{T}) = \nabla f(\boldsymbol{T}) : \dot{\boldsymbol{T}} = \nabla f(\boldsymbol{T}) : (\boldsymbol{A}(\boldsymbol{T})[\boldsymbol{D}] + \psi(\boldsymbol{T},\boldsymbol{D})\boldsymbol{B}(\boldsymbol{T})).$$

Introducing $\mu(\boldsymbol{T})=1+\nabla f(\boldsymbol{T}):\boldsymbol{B}(\boldsymbol{T})$, we get a more compact equation:

$$\dot{k} = \psi(\boldsymbol{T},\boldsymbol{D})\mu(\boldsymbol{T}) \qquad (5)$$

We say that hardening, softening or perfect plasticity takes place, in an interval $[t_0, t_1]$ in which the response is plastic, according to whether $\dot{k}$ $(=\nabla f(\boldsymbol{T}):\dot{\boldsymbol{T}})$ is positive, negative or null in $(t_0,t_1)$, which is determined by $\mu(\boldsymbol{T})$, since $\psi(\boldsymbol{T}(t):\boldsymbol{D}(t))>0$, $\forall t \in (t_0,t_1)$. Thus, the surface *S* defined by $\mu(\boldsymbol{T})=0$ divides the stress space into regions where either hardening or softening may happen. *S* will be called the limit surface.

An important conclusion is that, in plastic response, a stress path such that either $\nabla f(\boldsymbol{T}):\dot{\boldsymbol{T}} > 0$ or $\nabla f(\boldsymbol{T}):\dot{\boldsymbol{T}} < 0$ at all of its points cannot cross the limit surface. In section 4 we shall investigate what happens as stress paths of this kind approach *S*.

It is interesting to see the relation between cases II, III and IV (in the definition of material response) and $\nabla f(\boldsymbol{T}_0):\dot{\boldsymbol{T}}(t_0)$.

In case II: $\nabla f(\boldsymbol{T}_0):\dot{\boldsymbol{T}}(t_0) < 0$, and elastic response follows.

In case III: $\nabla f(\boldsymbol{T}_0):\dot{\boldsymbol{T}}(t_0) = 0$, and one or the other response takes place depending on whether $t \mapsto \psi(\boldsymbol{T}(t),\boldsymbol{D}(t))$ increases or not in a right-neighbourhood of $t_0$.

In case IV: $\nabla f(\boldsymbol{T}_0):\dot{\boldsymbol{T}}(t_0)$ is negative, zero or positive depending on $\mu(\boldsymbol{T})$ and plastic response follows.

Note that the definition of material response could not be established in terms of $\nabla f(\boldsymbol{T}_0):\dot{\boldsymbol{T}}(t_0)$, since, if it is negative and $f(\boldsymbol{T}_0)=k_0$, then the response may be elastic or plastic with softening in a right-neighbourhood of $t_0$. However, if it is positive and $f(\boldsymbol{T}_0)=k_0$, the response is certainly plastic with hardening.

Traditionally, $\nabla f(\boldsymbol{T}_0):\dot{\boldsymbol{T}}(t_0) = 0$ is associated to neutral loading. This terminology is not appropriate since plastic response may follow.

In view of what has been presented so far we can condense equations (2) and (3) in just one equation, if $\mu(\boldsymbol{T}) \neq 0$:

$$\overset{\circ}{\boldsymbol{T}} = A(\boldsymbol{T})[\boldsymbol{D}] + \frac{\dot{k}}{\mu(\boldsymbol{T})} B(\boldsymbol{T})$$

Recall that, in elastic response, $\dot{k}=0$ and, in plastic response, (5) holds.

## 4. FURTHER PROPERTIES OF THE LIMIT SURFACE

As explained in section 2, given the deformation gradient at each instant, the corresponding Cauchy stress, in sub-intervals, is the solution of a differential equation:

$$\dot{\boldsymbol{T}} = h(\boldsymbol{T},\boldsymbol{D}) - \boldsymbol{T}\boldsymbol{W} + \boldsymbol{W}\boldsymbol{T}, \qquad (6)$$

where $h(\boldsymbol{T},\boldsymbol{D})=A(\boldsymbol{T})[\boldsymbol{D}]$, if the response is elastic, and $h(\boldsymbol{T},\boldsymbol{D})=A(\boldsymbol{T})[\boldsymbol{D}]+\psi(\boldsymbol{T},\boldsymbol{D})B(\boldsymbol{T})$, if the response is plastic. In both cases $h$ is isotropic, as we can deduce from the considerations of objectivity done in the end of section 2.

It will be convenient to work with the equation for plastic response in a more compact form. For that purpose recall that, for $M$, $N \in Sym$, the tensor product of $M$ and $N$ is defined by $(M \otimes N)[S]=(S:N)M$, for every $S \in Sym$. If we introduce, for each $T$, the fourth-order tensor $C(T)=A(T)+B(T) \otimes A(T)^T[\nabla f(T)]$, the expression of the function $h$ for plastic response can be written as $h(T,D)=C(T)[D]$.

We pass now to the investigation of the equilibrium points of the autonomous differential equation $\dot{T} = h(T,D)$ obtained from (6) when $D$ is constant and $W=0$. We say that $T$ is an equilibrium point, corresponding to $D$, if $h(T,D)=0$.

In the following analysis, for each $T$, the fourth-order tensor $A(T)$ is assumed to be invertible.

Consider first the equation for elastic response: $A(T)[D]=0$. If $D \neq 0$, then there are no equilibrium points, since $A(T)$ is invertible for every $T$. In the trivial case $D=0$, every $T$ is an equilibrium point.

More interesting is the discussion concerning the equation for plastic response.

*Proposition 2*: In plastic response, $T$ is an equilibrium point if and only if $\mu(T)=0$ (*i.e.*, $T$ belongs to the limit surface $S$) and the corresponding values of $D$ are $D = -\lambda A(T)^{-1}[B(T)]$, $\lambda$ being any positive real number.

*Proof*: We begin by finding the kernel of the linear operator $C(T)$, for each $T$, denoted $KerC(T)$.

If $D \in KerC(T)$, then $C(T)[D]=0$ and, therefore:

$$D = - \psi(T, D)A(T)^{-1}[B(T)].$$

So, every element of $KerC(T)$ is a multiple of $A(T)^{-1}[B(T)]$.

Now note that $C(T)[\alpha A(T)^{-1}[B(T)]] = \alpha\mu(T)B(T)$. Hence, $\alpha A(T)^{-1}[B(T)] \in KerC(T)$ if and only if $\mu(T)=0$ or $\alpha=0$. Thus, the kernel of $C(T)$ is

a) $\{\mathbf{0}\}$, if $\mu(T)\neq 0$;

b) $\{\alpha A(T)^{-1}[B(T)] \,/\, \alpha \in R\}$, if $\mu(T)=0$.

Since we are seeking the equilibrium points and the corresponding values of $D$ for plastic response, only those elements of $KerC(T)$ for which $\psi(T,D)>0$ are relevant. Hence, we must have $\mu(T)=0$ and $\alpha<0$, because $\psi(T,\alpha A(T)^{-1}[B(T)]) = \alpha\, B(T):\nabla f(T) = -\alpha$. (*End of the proof.*)

Due to the isotropy of $\mu$, $T$ is an equilibrium point if and only if $QTQ^T$ is an equilibrium point too, for every $Q\in Orth$. Besides, $D$ corresponds to $T$ if and only if $QDQ^T$ corresponds to $QTQ^T$.

Therefore, $S$ is composed of equilibrium points in the sense discussed above. So, if at a certain instant $T$ belongs to $S$ and to the yield surface, and a motion is imposed with null spin and constant stretching equal to $D = -\lambda A(T)^{-1}[B(T)]$, $\lambda>0$, then the response is plastic, and the stress will remain unchanged (a state of perfect plasticity).

Although $S$ is composed of equilibrium points corresponding only to stretchings of the form given in proposition 2, it plays an important role for other motions as well. We will see next, in proposition 3, what happens to the stretching as the stress approaches $S$ in plastic response coming from the region of hardening ($\mu(T)>0$ and $\nabla f(T):\dot{T}>0$) or from the region of softening ($\mu(T)<0$ and $\nabla f(T):\dot{T}<0$). An interesting result about the ratio between the distortion rate, represented by $\|\hat{D}\|$ ($\hat{D}$ is the deviatoric (traceless) part of $D$), and the volumetric strain rate, $|trD|$, will be obtained.

In proposition 3 we will use the fact that (3) can be solved for $D$ if $\mu(T)\neq 0$:

$$D = A(T)^{-1}[\overset{\circ}{T}] - \frac{\nabla f(T):\dot{T}}{\mu(T)} A(T)^{-1}[B(T)] \qquad (7)$$

*Proposition 3*: Let $T_0$ be a point not on $S$ and $T_1$ a point on $S$. Further, consider a continuously differentiable curve $t \mapsto T(t)$ defined on an open interval containing [0,1], such that $T(0)=T_0$, $T(1)=T_1$ and either

a) $\mu(T(t))>0$, $\forall t \in [0,1)$, and $\nabla f(T(t)):\dot{T}(t) > 0$, $\forall t \in [0,1]$;

or

b) $\mu(T(t))<0$, $\forall t \in [0,1)$, and $\nabla f(T(t)):\dot{T}(t) < 0$, $\forall t \in [0,1]$.

To the stress path defined by the restriction of this curve to [0,1) there corresponds a plastic response with $D$ given by (7), whose norm tends to infinity as $t$ tends to 1. Moreover

$$\lim_{t \to 1^-} \frac{\|\hat{D}\|}{|trD|} = \sqrt{\frac{\|A^{-1}(T)[B(T)]\|^2}{(tr(A^{-1}(T)[B(T)]))^2} - \frac{1}{3}} \qquad (8)$$

*Proof*: This stress path corresponds to a plastic response with hardening, if (a) holds, or with softening, if (b) holds. In either case, as $t$ tends to 1, the norm of $D$ in (7) tends to infinity, since $\mu(T_1)=0$ and $\nabla f(T_1):\dot{T}(1) \neq 0$.

An easy computation, using (7), gives

$$\frac{\|\hat{D}\|^2}{(trD)^2} = \frac{\mu^2\|X\|^2 - 2\mu z X:Y + z^2\|Y\|^2}{\mu^2(trX)^2 - 2\mu z trX trY + z^2(trY)^2} - \frac{1}{3}$$

where $\mu=\mu(T)$, $X = A(T)^{-1}[\dot{T}]$, $Y = A(T)^{-1}[B(T)]$ and $z = \nabla f(T):\dot{T}$. Therefore, (8) follows.(*End of the proof.*)

If we add as an axiom the requirement that $tr(A(T)^{-1}[B(T)])=0$ on S, which imposes a restriction on the functions $A$ and $B$, then $S$ may called the critical state surface $S_c$, an important concept in soil mechanics (see, *e.g.*, [5]), because it is possible to keep the stress constant on $S_c$ while the material suffers an isochoric plastic deformation ($tr\boldsymbol{D}=0$). In this case, the limit (8), which represents the limit of the ratio between the distortion rate and the volumetric strain rate, becomes infinity, in agreement with what is observed in experiments with soils.

## 5. THE DECOMPOSITION OF THE STRETCHING INTO ELASTIC AND PLASTIC PARTS, THE HARDENING LAW AND THE NORMALITY CONDITION

The purpose of this section is to show that some elements of classical plasticity can be obtained in the present theory.

If $\boldsymbol{A}(\boldsymbol{T})$ is invertible, then, in elastic response, from (2), $\boldsymbol{D} = \boldsymbol{A}(\boldsymbol{T})^{-1}[\overset{\circ}{\boldsymbol{T}}]$ holds and, in plastic response, (7) holds

$$\boldsymbol{D} = \boldsymbol{A}(\boldsymbol{T})^{-1}[\overset{\circ}{\boldsymbol{T}}] - \frac{\nabla f(\boldsymbol{T}) : \dot{\boldsymbol{T}}}{\mu(\boldsymbol{T})} \boldsymbol{A}(\boldsymbol{T})^{-1}[\boldsymbol{B}(\boldsymbol{T})]$$

provided $\mu(\boldsymbol{T}) \neq 0$. The above expression suggests to interpret $\boldsymbol{D}$ as being the sum of elastic and plastic parts: $\boldsymbol{D}=\boldsymbol{D}^e+\boldsymbol{D}^p$. The elastic stretching is $\boldsymbol{D}^e = \boldsymbol{A}(\boldsymbol{T})^{-1}[\overset{\circ}{\boldsymbol{T}}]$, while the plastic stretching is:

$$\boldsymbol{D}^p = -\frac{\nabla f(\boldsymbol{T}) : \dot{\boldsymbol{T}}}{\mu(\boldsymbol{T})} \boldsymbol{A}(\boldsymbol{T})^{-1}[\boldsymbol{B}(\boldsymbol{T})] = -\psi(\boldsymbol{T},\boldsymbol{D})\boldsymbol{A}(\boldsymbol{T})^{-1}[\boldsymbol{B}(\boldsymbol{T})] \qquad (9)$$

if the response is plastic, and $\boldsymbol{D}^p=\boldsymbol{0}$, if the response is elastic.

Next step is to establish a relation between the evolution of $k$ and the plastic stretching: the so-called hardening rule. From (5) and (9) we get

$$\dot{k} = \frac{\|\boldsymbol{D}^p\|}{\|\boldsymbol{A}(\boldsymbol{T})^{-1}[\boldsymbol{B}(\boldsymbol{T})]\|}\mu(\boldsymbol{T}).  \quad (10)$$

Traditionally, the hardening rule is a primitive element of the theory whereas here it was derived. The above expression shows us which form the hardening rule must have as well as its relation with other functions of the theory. Note that $\dot{k}$ is given by an isotropic function of $\boldsymbol{T}$ and $\boldsymbol{D}^p$ that is positively homogeneous of degree one in $\boldsymbol{D}^p$.

We will now see how the normality condition, an important ingredient of classical plasticity, can be introduced in this theory. It suffices to choose $\boldsymbol{B}$ given by $\boldsymbol{B}(\boldsymbol{T})=\beta(\boldsymbol{T})\boldsymbol{A}(\boldsymbol{T})[\nabla f(\boldsymbol{T})]$, with $\beta:Sym \to R$, isotropic, $\beta(\boldsymbol{T})<0$, for all $\boldsymbol{T}$. Thus, from (9):

$$\boldsymbol{D}^p = -\psi(\boldsymbol{T},\boldsymbol{D})\beta(\boldsymbol{T})\nabla f(\boldsymbol{T}). \quad (11)$$

In this case it is said that the normality condition holds: at any instant of a plastic response $\boldsymbol{D}^p$ and $\nabla f(\boldsymbol{T})$ are linearly dependent and, therefore, $\boldsymbol{D}^p$ is normal to the yield surface at the point $\boldsymbol{T}$. Further, as $-\psi(\boldsymbol{T}, \boldsymbol{D})\beta(\boldsymbol{T})$ is positive, $\boldsymbol{D}^p$ and $\nabla f(\boldsymbol{T})$ point to the same side of the yield surface.

In the presence of the normality condition the power of the stressing on the plastic deformation, defined by $\dot{\boldsymbol{T}}:\boldsymbol{D}^p$, is positive in hardening and negative in softening since, by (11):

$$\dot{\boldsymbol{T}}:\boldsymbol{D}^p = -\psi(\boldsymbol{T},\boldsymbol{D})\beta(\boldsymbol{T})\dot{\boldsymbol{T}}:\nabla f(\boldsymbol{T})$$

Suppose now that $\boldsymbol{0} \in E(t_0)$, $\boldsymbol{T}_0 \in \partial E(t_0)$, that in $[t_0,t_1]$ the response is plastic, that $\boldsymbol{0} \in E(t)$, $\forall t \in [t_0,t_1]$, that the normality condition holds and the yield surface is convex. Then it is easy to conclude that the work of the stress on the plastic deformation in the interval $[t_0, t]$ per unit volume in the reference configuration (at $t_0$), defined by:

$$\int_{t_0}^{t}(\det\boldsymbol{F})\boldsymbol{T}:\boldsymbol{D}^p\,d\tau=-\int_{t_0}^{t}\det\boldsymbol{F}\psi(\boldsymbol{T},\boldsymbol{D})\beta(\boldsymbol{T})\boldsymbol{T}:\nabla f(\boldsymbol{T})\,d\tau$$

is positive ($\det\boldsymbol{F}$ is the determinant of $\boldsymbol{F}$).

## 6. FINAL REMARKS

It was shown that it is possible to construct a relatively simple theory of plasticity with a small number of elements (three constitutive functions) that describes the material response to any imposed motion, obeys the principle of material objectivity, represents hardening and softening, and yet is compatible with many characteristics of the classical theory. The hardening rule, for instance, was obtained as an outcome.

The definition of the material response deserved special attention. As the constitutive equations are differential equations, the material response must be specified in intervals, rather than at isolated instants.

A particularly interesting result is the existence of a limit surface, which divides the stress space into regions of hardening and softening. Interpreted in the light of the theory of ordinary differential equations, the limit surface is composed of equilibrium points of the equation for plastic response for certain values of the stretching tensor. This means that a state of perfect plasticity occurs on the limit surface for certain motions.

Finally, it should be noted that also elastic-perfectly plastic behaviour can be modelled within this framework. It suffices to impose that the initial yield surface and the limit surface $S$ coincide, which can be attained by an appropriate choice of the functions $f$ and $\boldsymbol{B}$. It follows that the yield surface is fixed and that, in plastic response, the stress remains on $S$.

E-mail: jjnader@usp.br

Address: Av. Prof. Almeida Prado, Departamento de Engenharia de Estruturas e Geotécnica, Escola Politécnica, Universidade de São Paulo, 05508-900, São Paulo, Brasil.